\documentclass[orivec,runningheads]{llncs}

\usepackage[T1]{fontenc}
\usepackage{graphicx}
\usepackage{booktabs}
\usepackage[misc]{ifsym}

\usepackage[version=4]{mhchem}

\usepackage{tikz}
\usetikzlibrary{petri,positioning}

\usepackage{amsmath}
\usepackage{amssymb}

\usepackage{interval}
\usepackage{siunitx}
\usepackage{booktabs}

\usepackage{nicematrix}
\NiceMatrixOptions{code-for-first-row=\color{blue},
                   code-for-last-row=\color{blue},
                   code-for-first-col=\color{blue},
                   code-for-last-col=\color{blue}}

\usepackage[hidelinks]{hyperref}
\begin{document}

\title{Machine Learning for Chemistry Reduction\\ in \ce{N2}--\ce{H2} Low-Temperature Plasmas}

\titlerunning{Machine Learning for Chemistry Reduction in \ce{N2}--\ce{H2} Plasmas}

\author{Diogo R. Ferreira\inst{1} \and Alexandre Lan\c{c}a\inst{1} \and Lu\'{i}s Lemos Alves\inst{2}}

\authorrunning{D. R. Ferreira \and A. Lan\c{c}a \and L. L. Alves}

\institute{Instituto Superior T\'{e}cnico, Universidade de Lisboa, 1049-001 Lisboa, Portugal \and Instituto de Plasmas e Fus\~{a}o Nuclear (IPFN), Instituto Superior T\'{e}cnico,\\ Universidade de Lisboa, 1049-001 Lisboa, Portugal\\ \email{diogo.ferreira@tecnico.ulisboa.pt}}

\maketitle

\begin{abstract}
Low-temperature plasmas are partially ionized gases, where ions and neutrals coexist in a highly reactive environment. This creates a rich chemistry, which is often difficult to understand in its full complexity. In this work, we develop a machine learning model to identify the most important reactions in a given chemical scheme. The training data are an initial distribution of species and a final distribution of species, which can be obtained from either experiments or simulations. The model is trained to provide a set of reaction weights, which become the basis for reducing the chemical scheme. The approach is applied to \ce{N2}--\ce{H2} plasmas, created by an electric discharge at low pressure, where the main goal is to produce \ce{NH3}. The interplay of multiple species, as well as of volume and surface reactions, make this chemistry especially challenging to understand. Reducing the chemical scheme via the proposed model helps identify the main chemical pathways.
\end{abstract}

\section{Introduction}
\label{sec:introduction}

The production of ammonia (\ce{NH3}) is critically important to a wide range of economic activities in the agricultural and industrial sectors. For example, ammonia is a key ingredient in the manufacturing of fertilizers, pharmaceuticals, plastics, textiles, and other chemicals. For more than a hundred years, the main industrial procedure for the production of ammonia has been the Haber-Bosch process~\cite{Erisman_2008}. However, this procedure requires temperatures in the range of 400 to 500 \textdegree{}C and pressures in the range of 150 to 300 atmospheres, which translate into considerable requirements in terms of energy consumption. These conditions are necessary in order to overcome the triple bond of nitrogen molecules (\ce{N2}), enabling them to react with hydrogen (\ce{H2}), via \ce{N2 + 3H2} $\rightarrow$ \ce{2NH3}.

Low-temperature plasmas have emerged as a more energy efficient and environmentally friendly process for ammonia synthesis~\cite{Carreon_2019}. In this case, the process begins by exciting and ionizing a gas mixture of \ce{N2} and \ce{H2}. Several ionization methods are available, but one of the simplest and most direct is to apply an electrical discharge~\cite{Chatain_2023}. This brings the gas mixture into a partially ionized state, containing electrons, ions, radicals, and neutral particles. In this plasma state, the triple bonds of \ce{N2} and the covalent bonds of \ce{H2} can be broken more effectively than in thermal processes, leading to the formation of highly reactive nitrogen and hydrogen species. A vast and complicated series of reactions then unfolds, eventually producing \ce{NH3}. One of the main goals of this work is to contribute to the understanding of this chain of reactions.

Studying the chemistry of ammonia production is important for additional reasons. Ammonia is known to be present in the atmosphere of the gas giants in our solar system, especially in Jupiter, Saturn, and Saturn's moon Titan~\cite{Encrenaz_1974}. Ammonia has also been detected in interstellar space and in exoplanets~\cite{Huang_2022}. The complex chemistry that takes place in a planetary atmosphere has prompted scientists to carry out laboratory experiments, including plasma discharges, to study the formation of organic molecules and even prebiotic molecules required for the origin of life~\cite{Chudjak_2021}. Understanding the production of ammonia in those environments, through lab experiments and computer-based simulations, is an integral part of those endeavors. Using machine learning, our goal is to identify the most important reactions in such complex chemical processes.

The paper is organized as follows. Section~\ref{sec:chemistry} provides a brief introduction to the chemistry of \ce{N2}--\ce{H2} plasmas. Section~\ref{sec:petrinets} explains how a chemical scheme containing many reactions can be represented as a Petri net. This representation can be translated into matrix form, and we use this matrix form to develop a machine learning model in Section~\ref{sec:model}. In Section~\ref{sec:reduction} we describe where the training data comes from, and how the model is trained. Finally, in Section~\ref{sec:analysis} we analyze the results and identify the main pathways for \ce{NH3} production.

\section{Chemistry of \ce{N2}--\ce{H2} Plasmas}
\label{sec:chemistry}

The basis of the chemistry of \ce{N2}--\ce{H2} low-temperature plasmas is fairly well documented in the literature~\cite{Jimenez-Redondo_2020}. It consists of over 200 reactions, including vibrational processes, electron impact processes, as well as several reactions involving heavy (neutral and charged) species. Since it is impractical (and unnecessary) to reproduce the chemical scheme here in full, we provide only a few examples of the type of reactions that it includes.

For vibrational processes, there are reactions such as:
\begin{itemize}
	
	\item Vibrational excitation or de-excitation, e.g. e + \ce{N2}(X,$v$) $\rightleftharpoons$ e + \ce{N2}(X,$w$), where an electron collides with a nitrogen molecule in a vibrational level $v$, and excites (or de-excites) it to a higher (or lower) vibrational level $w$. Similar reactions exist for hydrogen molecules.
	
	\item Vibrational energy transfer, e.g. \ce{N2}(X,$v$) + \ce{H2}(X,$w$) $\rightleftharpoons$ \ce{N2}(X,$v+1$) + \ce{H2}(X,$w-1$), where nitrogen and hydrogen molecules collide, and energy is transferred from one to the other through their vibrational levels.
	
\end{itemize}

For other electron impact processes, there are reactions such as:
\begin{itemize}
	
	\item Electronic excitation and de-excitation, e.g. e + \ce{N2}(X) $\rightleftharpoons$ e + \ce{N2(A)}, where the collision of an electron with a nitrogen molecule causes its excitation from the ground state (X) to a higher energy state (A), or vice-versa.
	
	\item Dissociation, e.g. e + \ce{H2}(X) $\rightarrow$ e + \ce{2H}, where the collision of an electron with a hydrogen molecule breaks its molecular bond and dissociates it into two hydrogen atoms.
	
	\item Ionization, e.g. e + \ce{N2}(X) $\rightarrow$ \ce{N2+} + 2e, where the collision of an electron with a nitrogen molecule ejects an electron from the molecule, thereby ionizing it.
	
	\item Attachment and detachment, e.g. e + \ce{H2} $\rightarrow$ \ce{H} + \ce{H-}, where the collision of an electron with a hydrogen molecule causes the dissociation of the molecule, and the attachment of the electron to one of the resulting hydrogen atoms, forming a hydride ion (\ce{H-}).
	
	\item Neutralization, e.g. e + \ce{N2+} $\rightarrow$ \ce{N} + \ce{N}, where the incoming electron recombines with the positively charged nitrogen ion, thereby neutralizing it, and, in addition, the excess energy of the newly formed molecular state leads to the breaking of the molecular bond between the two nitrogen atoms.
	
\end{itemize}

For heavy-species collisions, there are reactions such as:
\begin{itemize}
	
	\item Collisional excitation and de-excitation, possibly with dissociation, e.g. \ce{N2}(A) + \ce{H2} $\rightarrow$ \ce{N2}(X) + \ce{2H}, where a nitrogen molecule in an excited state transfers energy to a hydrogen molecule. The nitrogen molecule returns to its ground state, and the energy transferred breaks the molecular bond between the two hydrogen atoms.
	
	\item Ionization, e.g. \ce{N2}(A) + \ce{N2}($\mathrm{a'}$) $\rightarrow$ \ce{N2}(X) + \ce{N2+} + e, where there is an energy transfer between two nitrogen molecules in different excited states. One of them returns to the ground state, while the other receives sufficient energy to eject an electron and ionize.
	
	\item Neutral–neutral reactions, e.g. \ce{N(D)} + \ce{H2(X)} $\rightarrow$ \ce{H} + \ce{NH}, where an excited nitrogen atom transfers energy to a hydrogen molecule, enough to break its molecular bond. The nitrogen atom then forms a new bond with one of the free hydrogen atoms.
	
	\item Ion-molecule reactions, e.g. \ce{N2+} + \ce{H2} $\rightarrow$ \ce{N2H+} + \ce{H}, where the nitrogen ion breaks the bond of a hydrogen molecule, and a new bond is formed between the nitrogen ion and one of the hydrogen atoms.
	
	\item Ion-ion recombination, e.g. \ce{H-} + \ce{H2+} $\rightarrow$ \ce{H2} + \ce{H}, where a hydride ion (\ce{H-}) transfers an excess electron to neutralize the charge of the dihydrogen cation (\ce{H2+}), resulting in a hydrogen molecule and a hydrogen atom.
	
\end{itemize}

The reactions above illustrate the types of mechanisms at work, and these mechanisms apply not only to \ce{N2} and \ce{H2} but also to atoms, radicals and ions resulting from the various chemical reactions. The chemistry is therefore rich enough to create a multitude of species (namely \ce{H}, \ce{H+}, \ce{H2}, \ce{H2+}, \ce{H3+}, \ce{N}, \ce{N+}, \ce{N2}, \ce{N2+}, \ce{N3+}, \ce{N4+}, \ce{NH}, \ce{NH+}, \ce{NH2}, \ce{NH2+}, \ce{NH2-}, \ce{NH3}, \ce{NH3+}, \ce{NH4+}, \ce{N2H+}, \ce{N2H2}, \ce{N2H3}, \ce{N2H4}) and these species can be in different excited states, particularly electronic and vibrational. For simplicity, we will omit the details about the electronic/vibrational configuration of those species.


Besides the reactions that take place in the bulk of the plasma -- often referred to as \emph{volume} processes~\cite{Gordiets_1998a} -- there is another set of reactions that take place at the boundaries of the experiment (at the interface between the gas and its surrounding walls), which are referred to as \emph{surface} processes~\cite{Gordiets_1998b}. The plasma-surface interaction stimulates the reaction of some species (namely \ce{H}, \ce{N}, \ce{NH}, \ce{NH2}) with the surface, by some physical or chemical means. A species may stick to the surface (physically) or react with the surface (chemically). Subsequently, there may be reactions between species that have previously adhered to the wall (involving only surface species), and there may be reactions between species that have previously adhered to the wall and species that come from the bulk of the plasma (involving both surface and volume species). In some cases, the products of these reactions can return to the volume.

To distinguish between volume and surface species, we use the ``wall'' prefix for the latter. For example, \ce{NH2} refers to the volume species, but \ce{wall\_NH2} refers to the surface species. In addition, we represent vacant sites where species can physically attach to the surface as \ce{wall\_F_v}, and vacant sites where species can chemically bind to the surface as \ce{wall\_S_v}.

The surface chemistry can then be formulated as, for example:

\begin{itemize}
	
	\item \ce{NH(X)} + \ce{wall\_S_v} $\rightarrow$ \ce{wall\_NH(S)}, where \ce{NH} adheres chemically to the wall.
	
	\item \ce{NH(X)} + \ce{wall\_H(S)} $\rightarrow$ \ce{wall\_NH2(S)}, where an incoming \ce{NH} molecule reacts with a hydrogen atom that is chemically bound to the wall.
	
	\item \ce{wall\_H(F)} + \ce{wall\_NH(S)} $\rightarrow$ \ce{wall\_NH2(S)} + \ce{wall\_F_v}, where a hydrogen atom that was physically attached to the wall reacts with a \ce{NH} molecule that is chemically bound to the wall. The resulting \ce{NH2} is bound to the wall and the physical site that released the hydrogen atom becomes vacant.
	
	\item \ce{H2(X)} + \ce{wall\_NH(S)} $\rightarrow$ \ce{NH3(X)} + \ce{wall\_S_v}, where an incoming hydrogen molecule reacts with a \ce{NH} molecule that was chemically bound to the wall, releasing \ce{NH3} and leaving the chemical site vacant.

\end{itemize}

These examples illustrate the surface chemistry that needs to be taken into account, simultaneously with the reactions that take place in the plasma volume.

\section{Petri Net Modeling}
\label{sec:petrinets}

One way to model chemical reactions is through the use of Petri nets~\cite{Marwan_2011}. In essence, a Petri net is a graphical model with two types of nodes: \emph{places} and \emph{transitions}. Places connect to transitions, and transitions connect to places, through directed arcs. In addition, places have \emph{tokens}. When a transition \emph{fires}, it removes tokens from its input places, and adds tokens to its output places. By default, it removes one token from each of its input places, and adds one token to each of its output places. However, the number of tokens to be removed from (or added to) each place can be configured by labeling the arc connecting the place to (or from) the transition. For a transition to be \emph{enabled}, each of its input places must contain a sufficient number of tokens; otherwise, it cannot fire.

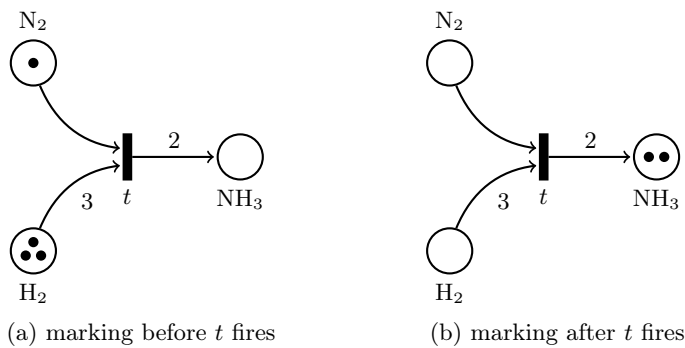
\begin{figure}[t]
	\centering
	\begin{tabular}{c@{\hskip 2cm}c}
		\begin{tikzpicture}[thick, every place/.style={thick,minimum size=6mm}, every transition/.style={fill,minimum width=1mm,minimum height=6mm}]
			
			\node[place,label=above:$\ce{N2}$,tokens=1] (N2) at (0,0) {};
			\node[place,label=below:$\ce{H2}$,tokens=3] (H2) at (0,-2.5) {};
			\node[place,label=below:$\ce{NH3}$,tokens=0] (NH3) at (2.75,-1.25) {};
			
			\node[transition,label=below:$t$] (t) at (1.25,-1.25) {};
			
			\draw (N2) edge[post, bend right] (t.120);
			\draw (H2) edge[post, bend left] node[auto,swap] {3} (t.-120);
			\draw (t) edge[post] node[auto] {2} (NH3);
			
		\end{tikzpicture} &
		\begin{tikzpicture}[thick, every place/.style={thick,minimum size=6mm}, every transition/.style={fill,minimum width=1mm,minimum height=6mm}]
			
			\node[place,label=above:$\ce{N2}$,tokens=0] (N2) at (0,0) {};
			\node[place,label=below:$\ce{H2}$,tokens=0] (H2) at (0,-2.5) {};
			\node[place,label=below:$\ce{NH3}$,tokens=2] (NH3) at (2.75,-1.25) {};
			
			\node[transition,label=below:$t$] (t) at (1.25,-1.25) {};
			
			\draw (N2) edge[post, bend right] (t.120);
			\draw (H2) edge[post, bend left] node[auto,swap] {3} (t.-120);
			\draw (t) edge[post] node[auto] {2} (NH3);
			
		\end{tikzpicture} \\
		(a) marking before $t$ fires & (b) marking after $t$ fires
	\end{tabular}
	\caption{A chemical reaction modeled as a Petri net transition $t$.}
	\label{fig:petrinet:reaction}
\end{figure}

Figure~\ref{fig:petrinet:reaction} shows an example of how a chemical reaction can be represented as a Petri net transition, with the reactants as input places, and the products as output places. For illustrative purposes, we use the Haber-Bosch reaction \ce{N2 + 3H2} $\rightarrow$ \ce{2NH3}. In Fig.~\ref{fig:petrinet:reaction}(a), the transition $t$ is enabled by having a sufficient number of tokens in each of its input places; this is called the initial \emph{marking}. When $t$ fires, one token is subtracted from place \ce{N2}, three tokens are subtracted from place \ce{H2}, and two tokens are added to place \ce{NH3}. The resulting marking is shown in Fig.~\ref{fig:petrinet:reaction}(b). In this case, $t$ is no longer enabled but, in a general case, the presence of tokens in \ce{NH3} could enable subsequent transitions.

The same principles can be applied to model chemical reaction networks~\cite{Koch_2010}. The idea is perhaps better explained with an example. For this purpose, we consider a set of reactions that is by no means a complete description of the \ce{N2}--\ce{H2} plasma chemistry, but nevertheless includes some of the mechanisms that are present in such chemistry, with the mere purpose of illustrating an over-simplified ammonia production process:
\begin{itemize}
	
	\item \ce{N2} $\rightleftharpoons$ \ce{2N}, which can be regarded as two separate (forward and backward) reactions, to be represented as transitions $t_1$ and $t_2$.
	
	\item \ce{N2} + \ce{H2} $\rightarrow$ \ce{N2} + \ce{2H}, to be represented as transition $t_3$.
	
	\item \ce{2H} $\rightarrow$ \ce{H2}, to be represented as transition $t_4$.
	
	\item \ce{N} + \ce{H} $\rightarrow$ \ce{NH}, to be represented as transition $t_5$.
	
	\item \ce{NH} + \ce{H} $\rightarrow$ \ce{NH2}, to be represented as transition $t_6$.
	
	\item \ce{NH2} + \ce{H} $\rightarrow$ \ce{NH3}, to be represented as transition $t_7$.
	
	\item \ce{NH3} + \ce{N} $\rightarrow$ \ce{NH2} + \ce{NH}, to be represented as transition $t_8$.
	
\end{itemize}

(In this example, the list of reactions ignores the presence of electrons, ions and  wall species, and accordingly does not consider ionization reactions.)

\begin{figure}[t]
	\centering
	\begin{tikzpicture}[thick, every place/.style={thick,minimum size=6mm}, every transition/.style={fill,minimum width=1mm,minimum height=6mm}]
		
		\node[place,label=left:$\ce{N2}$,tokens=2] (N2) at (0,0) {};
		\node[place,label=above:$\ce{N}$] (N) at (2.5,1.25) {};
		\node[place,label=left:$\ce{H2}$,tokens=2] (H2) at (0,-2.5) {};
		\node[place,label=93:$\ce{H}$] (H) at (2.5,-1.25) {};
		\node[place,label=right:$\ce{NH}$] (NH) at (5,0) {};
		\node[place,label=right:$\ce{NH2}$] (NH2) at (7.5,-1.25) {};
		\node[place,label=right:$\ce{NH3}$] (NH3) at (10,-2.5) {};
		
		\node[transition,label=below:$t_1$] (t1) at (1.25,1.25) {};
		\node[transition,label=below:$t_2$] (t2) at (1.5,0) {};
		\node[transition,label=below:$t_3$] (t3) at (1.25,-1.25) {};
		\node[transition,label=below:$t_4$] (t4) at (1.5,-2.5) {};
		\node[transition,label=below:$t_5$] (t5) at (3.75,0) {};
		\node[transition,label=below:$t_6$] (t6) at (6.25,-1.25) {};
		\node[transition,label=below:$t_7$] (t7) at (8.75,-2.5) {};
		\node[transition,label=below:$t_8$] (t8) at (6.5,1.25) {};
		
		\draw (N2) edge[post, bend left] (t1);
		\draw (t1) edge[post] node[auto,swap] {2} (N);
		\draw (N) edge[post, bend left] node[auto] {2} (t2);
		\draw (t2) edge[post] (N2);
		\draw (N2) edge[post, bend right] (t3.120);
		\draw (H2) edge[post, bend left] (t3.-120);
		\draw (t3) edge[post, bend right] (N2);
		\draw (t3) edge[post] node[auto,swap] {2} (H);
		\draw (H) edge[post, bend left] node[auto] {2} (t4);
		\draw (t4) edge[post] (H2);
		\draw (N) edge[post, bend right] (t5.120);
		\draw (H) edge[post, bend left] (t5.-120);
		\draw (t5) edge[post] (NH);
		\draw (NH) edge[post, bend right] (t6.120);
		\draw (H) edge[post] (t6);
		\draw (t6) edge[post] (NH2);
		\draw (NH2) edge[post, bend right] (t7.120);
		\draw (H) edge[post, bend right=12.5] (t7.-120);
		\draw (t7) edge[post] (NH3);
		\draw (N) edge[post] (t8);
		\draw (NH3) edge[post, bend right] (t8);
		\draw (t8.-120) edge[post, bend right] (NH);
		\draw (t8.-60) edge[post, bend left] (NH2);
		
	\end{tikzpicture}
	\caption{A chemical reaction network modeled as a Petri net.}
	\label{fig:petrinet:network}
\end{figure}
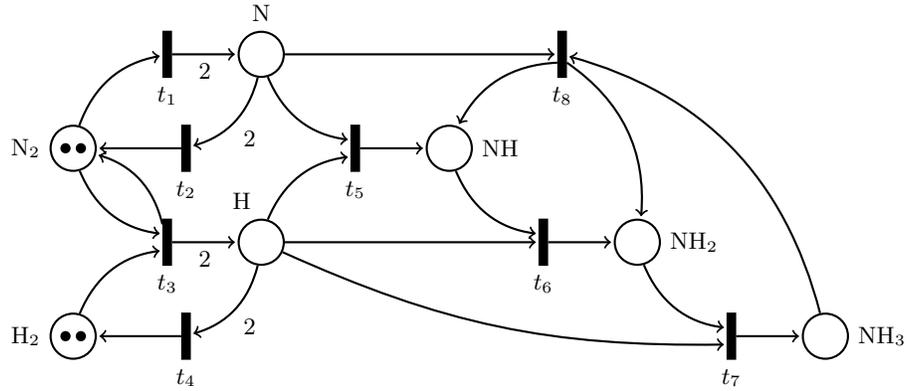

Figure~\ref{fig:petrinet:network} shows the corresponding Petri net. One can imagine that by placing a generous amount of tokens in places \ce{N2} and \ce{H2}, such initial marking could enable a series of transition firings until, eventually, some amounts of \ce{NH}, \ce{NH2} and \ce{NH3} are produced. Due to the cyclic chaining of some of these reactions, such firings could go on indefinitely until the experiment is interrupted, or until all the hydrogen and/or nitrogen tokens are consumed.

Mathematically, the structure of a Petri net can be represented as a transition matrix, where each element specifies the number of tokens that are subtracted from (or added to) a given place by a given transition. For the example in Fig.~\ref{fig:petrinet:network}, the transition matrix is:

\begin{equation}
	A = 
	\begin{bNiceArray}{rrrrrrrr}[margin,first-row,last-col,columns-width=5mm]
		t_1 & t_2 & t_3 & t_4 & t_5 & t_6 & t_7 & t_8 &          \\
		0  &  0  &  2  & -2  & -1  & -1  & -1  &  0  & \ce{H}   \\
		0  &  0  & -1  &  1  &  0  &  0  &  0  &  0  & \ce{H2}  \\
		2  & -2  &  0  &  0  & -1  &  0  &  0  & -1  & \ce{N}   \\
		-1  &  1  &  0  &  0  &  0  &  0  &  0  &  0  & \ce{N2}  \\
		0  &  0  &  0  &  0  &  1  & -1  &  0  &  1  & \ce{NH}  \\
		0  &  0  &  0  &  0  &  0  &  1  & -1  &  1  & \ce{NH2} \\
		0  &  0  &  0  &  0  &  0  &  0  &  1  & -1  & \ce{NH3} \\
	\end{bNiceArray}
	\label{eq:A}
\end{equation}

Let us suppose that, initially, there are two tokens in \ce{N2} and two tokens in \ce{H2}. This initial marking of the Petri net can be represented as a vector:
\begin{equation}
	b = 
	\begin{bNiceArray}{c}[margin,last-col]
		0  & \ce{H}   \\
		2  & \ce{H2}  \\
		0  & \ce{N}   \\
		2  & \ce{N2}  \\
		0  & \ce{NH}  \\
		0  & \ce{NH2} \\
		0  & \ce{NH3} \\
	\end{bNiceArray}
	\label{eq:b}
\end{equation}

From this initial marking, let us suppose that $t_1$ fires once, $t_3$ fires twice, and $t_5$, $t_6$, $t_7$ each fire once. This can be represented as a firing vector:
\begin{equation}
	x = 
	\begin{bNiceArray}{c}[margin,last-col]
		1  & t_1 \\
		0  & t_2 \\
		2  & t_3 \\
		0  & t_4 \\
		1  & t_5 \\
		1  & t_6 \\
		1  & t_7 \\
		0  & t_8 \\
	\end{bNiceArray}
	\label{eq:x}
\end{equation}

Then the final marking of the Petri net can be calculated as $Ax + b$:
\begin{equation}
	y = Ax + b = 
	\begin{bNiceArray}{c}[margin,last-col]
		1  & \ce{H}   \\
		0  & \ce{H2}  \\
		1  & \ce{N}   \\
		1  & \ce{N2}  \\
		0  & \ce{NH}  \\
		0  & \ce{NH2} \\
		1  & \ce{NH3} \\
	\end{bNiceArray}
	\label{eq:y}
\end{equation}

(The interested reader might want to check that this is indeed the case by imagining the movement of tokens in the Petri net of Fig.~\ref{fig:petrinet:network}.)

In this formulation, the most important problem for this work is how to find $x$ given $A$, $b$ and $y$. In general, the chemical scheme (i.e. the set of possible reactions), represented as matrix $A$, is known. The initial amounts of hydrogen and nitrogen that are provided at the beginning of the experiment or simulation, represented as vector $b$, are also known. The final amounts of species at the end of the experiment or simulation, represented as vector $y$, can be measured or calculated. Therefore, what is left to know is the vector $x$, which is proxy for the reaction rates, i.e.~the number of times those reactions occur per unit time and unit volume (or surface). In this work, we will interpret $x$ as a measure of weight or importance of each reaction in the chemical scheme.

\section{Machine Learning Model}
\label{sec:model}

As we have seen in the previous section, $A$ is an $n\!\times\!m$ matrix, where $n$ is the number of species, and $m$ is the number of reactions. In pratice, depending on the chemical scheme, the number of reactions $m$ can be (much) larger than the number of species $n$. This means that $Ax + b = y$ is an under-determined system. While there might be many solutions for $x$, a possible approach would be to find an approximation $\tilde{x}$ (and its corresponding prediction $\tilde{y} = A\tilde{x} + b$) that minimizes the residual $\lVert \tilde{y} - y \rVert$ and has the smallest norm $\lVert \tilde{x} \rVert$.

In our case, there are additional restrictions to be imposed on the model:
\begin{itemize}
	
	\item The first restriction is that every element of the solution (or approximation) $\tilde{x}$ should be non-negative, i.e.~$\forall_i: \tilde{x}_i \geq 0$. This means that negative transition firings (or, in a chemical sense, negative reaction rates) are disallowed.
	
	\item The second restriction has to do with how the experimental/calculated data about the initial and final amounts (i.e.~the density) of species are typically provided. While absolute values for the density of species (in m$^{-3}$ for volume species, and in m$^{-2}$ for surface species) are usually available, it is quite common to describe an experiment as comprising certain percentages of molecular hydrogen and nitrogen as input (e.g. 5\% of \ce{H2} and 95\% of \ce{N2}) and producing a certain \emph{distribution} of species as output (also expressed in percentages). As a result, this means that the model should provide a normalized output $\tilde{y}$. However, the distributions of volume species and surface species should be normalized separately, hence part of $\tilde{y}$ (the partition $\tilde{y}_{i \in \mathcal{V}}$ that corresponds to volume species) is normalized to 1, and another part of $\tilde{y}$ (the partition $\tilde{y}_{i \in \mathcal{S}}$ that corresponds to surface species) is also normalized to 1. In other words:
	\begin{equation}
		\begin{cases}
			\sum_{i \in \mathcal{V}} \tilde{y}_i = 1 & \text{for volume species in } \mathcal{V}=\{\ce{H}, \ce{H2}, \ce{N}, \ce{N2}, ...\} \\
			\sum_{i \in \mathcal{S}} \tilde{y}_i = 1 & \text{for surface species in } \mathcal{S}=\{\ce{wall\_H}, \ce{wall\_N}, ...\}
		\end{cases}
		\label{eq:normalization}
	\end{equation}
	
\end{itemize}

The non-negativity restriction on $\tilde{x}$ and the normalization requirements on $\tilde{y}$ can be implemented via standard machine learning constructs, one of them being the \emph{rectified linear unit} (ReLU) and the other being the \emph{softmax} function. Both functions are extensively used in the field of machine learning~\cite{Asadi_2020}.

In particular, ReLU is an element-wise function that can be defined as:
\begin{equation}
	\mathtt{ReLU}(\tilde{x}) = \left[\ldots,\mathtt{ReLU}(\tilde{x}_i),\ldots\right]^\top = \left[\ldots,\max(0, \tilde{x}_i),\ldots\right]^\top
	\label{eq:relu}
\end{equation}

On the other hand, softmax is a normalization function defined as:
\begin{equation}
	\mathtt{softmax}(\tilde{y}) = \left[\ldots,\mathtt{softmax}(\tilde{y}_i),\ldots\right]^\top = \left[\ldots,\frac{e^{\tilde{y}_i}}{\sum_j e^{\tilde{y}_j}},\ldots\right]^\top
	\label{eq:softmax}
\end{equation}

The purpose of using these functions is to allow $\tilde{x}$ and $\tilde{y}$ to remain unconstrained, while enforcing the non-negativity of $\tilde{x}$ and the normalization of $\tilde{y}$ by applying ReLU and softmax to them, respectively.

Since softmax is applied separately on volume and surface species, the model can be expressed as:
\begin{equation}
	\begin{cases}
		\mathtt{softmax}\left((A \cdot \mathtt{ReLU}(\tilde{x}) + b)_{i \in \mathcal{V}}\right) = \tilde{y}_{i \in \mathcal{V}} \\
		\mathtt{softmax}\left((A \cdot \mathtt{ReLU}(\tilde{x}) + b)_{i \in \mathcal{S}}\right) = \tilde{y}_{i \in \mathcal{S}} \\
	\end{cases}
	\label{eq:model}
\end{equation}

In the model above, a guess for $\tilde{x}$ will produce predictions for $\tilde{y}_{i \in \mathcal{V}}$ and $\tilde{y}_{i \in \mathcal{S}}$. The goal is to find the best guess $\tilde{x}$, i.e.~the guess $\tilde{x}$ for which the predictions $\tilde{y}_{i \in \mathcal{V}}$ and $\tilde{y}_{i \in \mathcal{S}}$ best approximate, in some sense, the true $y_{i \in \mathcal{V}}$ and $y_{i \in \mathcal{S}}$.

Now, if the true $y_{i \in \mathcal{V}}$ and the prediction $\tilde{y}_{i \in \mathcal{V}}$ are both normalized (the same applies to $y_{i \in \mathcal{S}}$ and $\tilde{y}_{i \in \mathcal{S}}$), they can be reinterpreted as probability distributions. This provides the opportunity for using a loss function based on statistical measures, such as the Kullback–Leibler (KL) divergence, to quantify the difference between a true probability distribution $y_{i \in \mathcal{V}}$ and a predicted probability distribution $\tilde{y}_{i \in \mathcal{V}}$ (and, similarly, between $y_{i \in \mathcal{S}}$ and $\tilde{y}_{i \in \mathcal{S}}$).

Specifically, the KL divergence measures how much information is lost when a true distribution $p$ is approximated by a predicted distribution $q$. In this case, the KL divergence is defined as:
\begin{equation}
	D_{\text{KL}}(p \parallel q) = \sum_i p_i \log\left(\frac{p_i}{q_i}\right)
	\label{eq:kldiv}
\end{equation}
where $p$ and $q$ and vectors of probabilities over the same outcomes.

In our case, the loss function is defined as:
\begin{equation}
	\begin{split}
		L(\tilde{x}) & = D_{\text{KL}}(y_{i \in \mathcal{V}} \parallel \tilde{y}_{i \in \mathcal{V}}) + 
		D_{\text{KL}}(y_{i \in \mathcal{S}} \parallel \tilde{y}_{i \in \mathcal{S}}) \\
		& = \sum_{i \in \mathcal{V}} y_i \log\left(\frac{y_i}{\tilde{y}_i}\right) +
		\sum_{i \in \mathcal{S}} y_i \log\left(\frac{y_i}{\tilde{y}_i}\right)
		\label{eq:loss}
	\end{split}
\end{equation}

The goal is to find $\tilde{x}$ that minimizes $L(\tilde{x})$. For this purpose, the gradient of $L(\tilde{x})$, denoted as $\nabla L(\tilde{x})$, provides the direction in which the loss increases. Since the objective is to minimize the loss, we update $\tilde{x}$ iteratively via:
\begin{equation}
	\tilde{x}^{(k+1)} = \tilde{x}^{(k)} - \eta \nabla L(\tilde{x}^{(k)})
	\label{eq:learning}
\end{equation}
where $\eta$ is a positive learning rate. In other words, the model is trained by \emph{gradient descent}, or variants thereof~\cite{Ruder_2017}.

At the end of the training process, $\tilde{x}$ will provide the weights assigned to the reactions in the chemical scheme. Any negative weight found in $\tilde{x}$, by force of the ReLU function, will be turned into zero, and therefore the corresponding reaction does not contribute to the result $\tilde{y}$. In fact, all those reactions for which $\tilde{x}_i \leq 0$ can be considered as candidates for removal from the chemical scheme, and this becomes the basis for chemistry reduction in this work.

\section{Chemical Scheme Reduction}
\label{sec:reduction}

In~\cite{Chatain_2023}, the authors provide an experimental study of \ce{N2}--\ce{H2} discharge plasmas, based on a low-pressure (50--500 Pa), high-voltage (1--3 kV), direct-current (10-40 mA) setup. The gas mixture, containing 0--5\% \ce{H2}, is pumped into a 23 cm-long tube with a 2 cm diameter, and mass spectrometry is used to measure the formation of neutral species (especially \ce{NH3}) and many ion species, such as \ce{H+}, \ce{H2+}, \ce{H3+}, \ce{N+}, \ce{N2+}, \ce{N3+}, \ce{N4+}, \ce{NH+}, \ce{NH2+}, \ce{NH3+}, \ce{NH4+}, and \ce{N2H+}. The formation of \ce{NH3} (and of positive ions in the form \ce{NH_x+}) was found to be correlated with current and pressure.

An alternative way to study the chemistry of low-temperature plasmas is through numerical modeling, using computer-based simulation tools, such as the LisbOn KInetics (LoKI) code~\cite{Tejero-del-Caz_2019,Guerra_2019}. LoKI comprises two modules:
\begin{itemize}
	
	\item The first module is a solver for the electron Boltzmann equation that provides the statistical distribution of electrons according to their kinetic energy, i.e.~the so-called \emph{electron energy distribution function} (often abbreviated as \emph{eedf}). Knowing the \emph{eedf} is crucial to estimate the rate coefficients at which electrons excite, dissociate and ionize atoms and molecules in the gas, in order to study the overall dynamics of the system.
	
	\item In a second module, LoKI handles the plasma chemistry by determining the reaction rates for a provided chemical scheme. The chemical scheme is specified in a configuration file that contains hundreds of reactions collected from the literature. A comprehensive list is provided in~\cite{Jimenez-Redondo_2020}. The rate coefficients of some reactions are specified in the chemical scheme, while others are determined based on the \emph{eedf} calculated by the first module. The reaction rates are calculated by multiplying the corresponding rate coefficients by the densities of the reactants intervening in each reaction. By balancing the creation and destruction rates for the different reactions in the chemical scheme, LoKI computes the final densities for all the species in the plasma.
	
\end{itemize}

LoKI receives the initial densities of species as input fractions in a configuration file (e.g.~5\% of \ce{H2} expressed as 0.05, and 95\% of \ce{N2} expressed as 0.95, for a pure \ce{N2}--\ce{H2} mixture similar to  experimental conditions). Then it computes and saves the final densities in an output file. These final densities can be converted to fractions, so that input and output fractions can be directly compared.

\begin{table}[!b]
	\centering
	\caption{Input and output fractions for a LoKI simulation.}
	\label{tab:fractions}
	\setlength{\tabcolsep}{12pt}
	\begin{tabular}{lll}
		\toprule
		Species &  Input fraction & Output fraction \\
		\midrule
		\ce{H}			&	0.045	&	\num{4.076311e-02}	\\
		\ce{H+}			&	0		&	\num{1.521387e-11}	\\
		\ce{H2}			&	0.005	&	\num{2.821790e-02}	\\
		\ce{H2+}		&	0		&	\num{1.232711e-12}	\\
		\ce{H3+}		&	0		&	\num{1.350015e-10}	\\
		\ce{N}			&	0		&	\num{1.385849e-02}	\\
		\ce{N+}			&	0		&	\num{1.149678e-12}	\\
		\ce{N2}			&	0.95	&	\num{9.171353e-01}	\\
		\ce{N2+}		&	0		&	\num{1.998340e-11}	\\
		\ce{N2H+}		&	0		&	\num{2.301581e-07}	\\
		\ce{N2H2}		&	0		&	\num{1.186624e-09}	\\
		\ce{N2H3}		&	0		&	\num{1.913678e-11}	\\
		\ce{N2H4}		&	0		&	\num{2.236752e-15}	\\
		\ce{N3+}		&	0		&	\num{8.794668e-11}	\\
		\ce{N4+}		&	0		&	\num{1.844171e-13}	\\
		\ce{NH}			&	0		&	\num{1.477812e-06}	\\
		\ce{NH+}		&	0		&	\num{1.616889e-12}	\\
		\ce{NH2}		&	0		&	\num{1.188158e-08}	\\
		\ce{NH2+}		&	0		&	\num{1.287649e-10}	\\
		\ce{NH2-}		&	0		&	\num{6.742121e-09}	\\
		\ce{NH3}		&	0		&	\num{2.347209e-05}	\\
		\ce{NH3+}		&	0		&	\num{1.444042e-09}	\\
		\ce{NH4+}		&	0		&	\num{2.753958e-08}	\\
		\ce{wall\_F_v}	&	0.9976	&	\num{9.974761e-01}	\\
		\ce{wall\_H}	&	0.002	&	\num{2.313191e-03}	\\
		\ce{wall\_N}	&	0.0002	&	\num{1.184274e-04}	\\
		\ce{wall\_NH}	&	0.0002	&	\num{9.229275e-05}	\\
		\ce{wall\_NH2}	&	0		&	\num{8.429000e-09}	\\
		\ce{wall\_S_v}	&	0		&	\num{1.202643e-08}	\\
		\bottomrule
	\end{tabular}
\end{table}

Table~\ref{tab:fractions} shows an example of the input and output fractions for a LoKI simulation. For simplicity, we abstract from vibrational levels and electronic states (e.g.~\ce{N2} represents all \ce{N2}($\Lambda$,$v$) species regardless of vibrational level $v$ and electronic state $\Lambda$). Therefore, the input/output fractions of each species include all the electronic and vibrational populations, if applicable. More importantly, the table shows volume species first, and surface species at the end. As can be seen, the input fractions are normalized separately, i.e.~the fractions for volume species sum up to 1, and the fractions for surface species also sum up to 1; the same applies to the output fractions.\footnote{The input fractions may be different for a simulation and for a lab experiment. For example, simulations usually define not only the fractions of molecular species (e.g. \ce{H2} and \ce{N2}) but also of atomic species (e.g.~\ce{H} and \ce{N}) and wall species (e.g.~\ce{wall\_H}, \ce{wall\_N}, etc.). On the other hand, in a lab experiment, the input for the gas mixture is limited to the actual fraction of gases available in commercial bottles.}

With the previous information, it is possible to apply the model described in Section~\ref{sec:model}, which includes matrix $A$ and vectors $b$ and $y$:
\begin{itemize}
	
	\item Matrix $A$ is built from the reactions in the chemical scheme, by following the same approach that led to Eq.~\eqref{eq:A}. Basically, matrix $A$ contains the stoichiometric coefficients of each reaction, with reactants being represented as negative values, and products being represented as positive values. Converting the chemical scheme to $A$ results in a matrix of size $29\!\times\!160$, where there are 160 reactions involving the 29 species in Table~\ref{tab:fractions}. 
	
	\item  As for $b$ and $y$, these vectors correspond, respectively, to the input and output fractions presented in Table~\ref{tab:fractions}. The conversion of those fractions to $b$ and $y$ results in two vectors of size $29\!\times\!1$.
	
\end{itemize}

In the framework of $Ax + b = y$, the goal would be to find the solution $x$ that explains how much each reaction contributes to transform the input fractions $b$ into the output fractions $y$, according to the chemical scheme $A$. In this context, $x$ will be a vector of size $160\!\times\!1$ with a weight for each reaction.

However, in Section~\ref{sec:model} we have seen that such system in under-determined, as shown here by the fact that we have 29 equations for 160 unknowns. Besides, there are certain restrictions to be imposed, namely the non-negativity of any approximation $\tilde{x}$, and the normalization of any prediction $\tilde{y}$.

This has led us to an approach in the form $\mathtt{softmax}(A \cdot \mathtt{ReLU}(\tilde{x}) + b) = \tilde{y}$, where the objective is to find an approximate solution $\tilde{x}$ via gradient descent, based on the model in Eq.~\eqref{eq:model}, the loss function in Eq.~\eqref{eq:loss}, and a learning rule similar to Eq.~\eqref{eq:learning}. In this work, we use the popular Adam optimizer~\cite{Kingma_2014} with a small learning rate, i.e.~$\eta = 10^{-4}$.

\begin{figure}[p]
	\centering
	\includegraphics[width=0.8\textwidth]{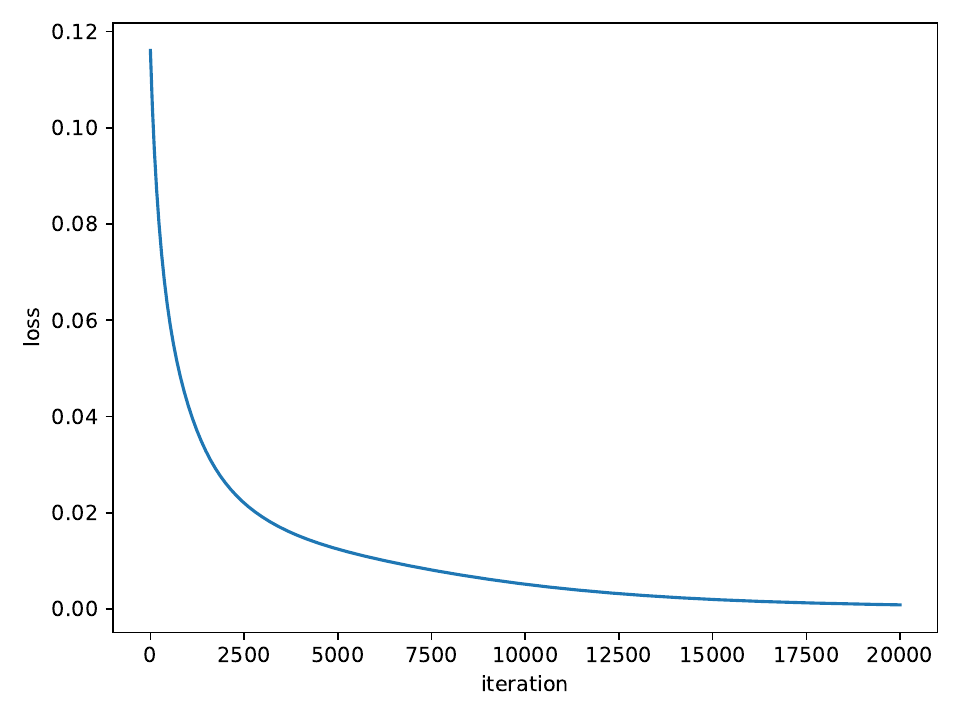}
	\caption{Loss curve obtained during training.}
	\label{fig:training:loss}
\end{figure}

Figure~\ref{fig:training:loss} shows the evolution of the loss values across training. The loss (i.e.~the KL divergence) decreases sharply during the first few thousand iterations, and then continues to improve, although at a much slower pace, until it eventually converges to a value on the order of $10^{-6}$.

\begin{figure}[p]
	\centering
	\includegraphics[width=\textwidth]{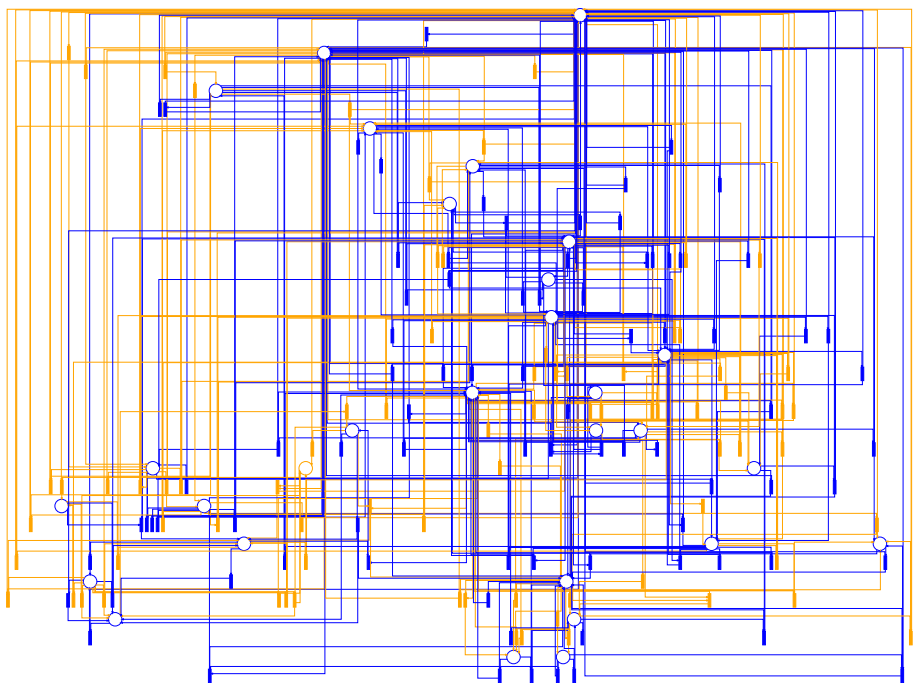}
	\caption{Petri net representation of the chemical scheme, with reactions to be kept (in blue), and reactions to be removed (in orange) [colors available in the online version].}
	\label{fig:scheme:petrinet}
\end{figure}

At the end of the training process, we extract $\tilde{x}$ and analyze the weights assigned to reactions. Reactions with a positive weight will be kept in the chemical scheme, as they are found to play a role in transforming the input fractions into the output fractions. On the other hand, reactions with zero or negative weight are considered as candidates for removal from the chemical scheme since, by force of the ReLU function, their contribution will be zero.

Figure~\ref{fig:scheme:petrinet} shows a Petri net representation of the chemical scheme, where the reactions with positive weight (these ones that will be kept) are shown in blue, and the reactions with zero or negative weight (the ones that will be considered for removal) are shown in orange. From the 160 reactions in the original chemical scheme, only about 60 reactions (less than 40\%) are guaranteed to survive, with the rest becoming candidates for removal. This highlights the extent to which the chemical scheme can be potentially reduced.

\section{Analysis of the Results}
\label{sec:analysis}

\begin{figure}[p]
	\centering
	\includegraphics[width=\textwidth]{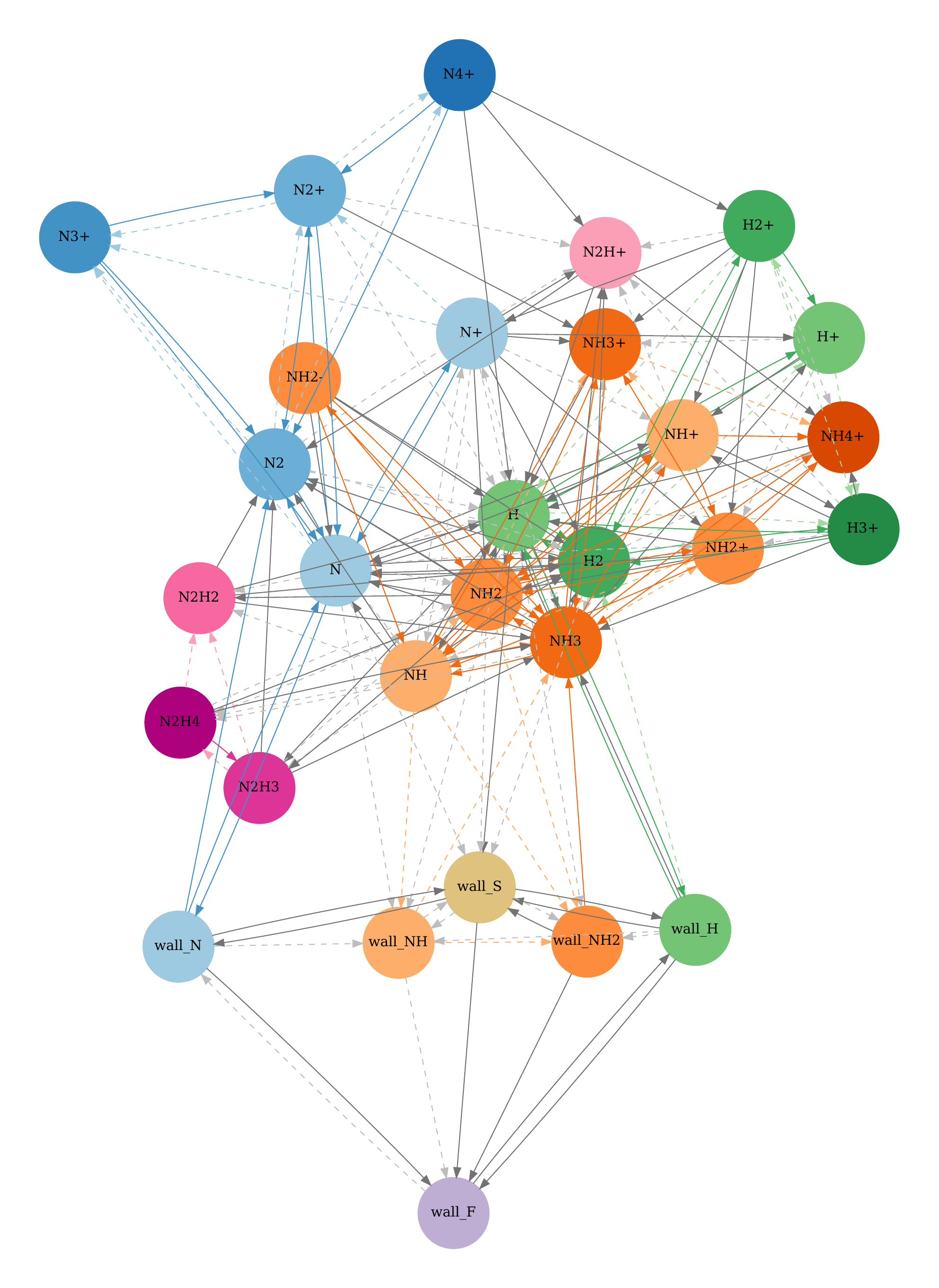}
	\caption{Graph representation of the chemical scheme, with reactions to be kept (solid lines), and reactions to be removed (dashed lines) [colors available in the online version].}
	\label{fig:scheme:graph}
\end{figure}

An analysis of the reduced chemical scheme, based on the machine learning model described in Section~\ref{sec:model}, reveals that there are several possible mechanisms for the production of \ce{NH3}:
\begin{itemize}
	
	\item One of those mechanisms is: \ce{wall\_H} + \ce{wall\_NH2} $\rightarrow$ \ce{NH3} (simplified form). In this case, a hydrogen atom attached to the wall reacts with a \ce{NH2} molecule bound to the wall, releasing \ce{NH3} into the plasma volume. This requires the availability of \ce{wall\_NH2}, which can be produced either by the surface chemistry (\ce{wall\_H} + \ce{wall\_NH} $\rightarrow$ \ce{wall\_NH2}) or may come from the volume chemistry (\ce{NH2} $\rightarrow$ \ce{wall\_NH2}). The former depends on the availability of \ce{wall\_NH}, which can be produced via similar pathways (either from \ce{wall\_H} and \ce{wall\_N}, or from the volume species \ce{NH} binding to the wall); the latter depends on the volume species \ce{NH2}, which can be produced in a number of ways, usually involving neutralization of \ce{NH2+} or \ce{NH2-}. These ion species can be traced back to a multitude of volume reactions.
	
	\item Another mechanism is \ce{H} + \ce{wall\_NH2} $\rightarrow$ \ce{NH3} (simplified form), where a hydrogen atom reacts with a \ce{NH2} molecule bound to the surface, releasing \ce{NH3} into the plasma volume. Much of the same considerations apply here as well, and this serves to illustrate how the surface chemistry can play an important role in the production of \ce{NH3}. Once \ce{H} and \ce{N} bind to the wall, they can react chemically to produce \ce{wall\_NH}, \ce{wall\_NH2}, and eventually \ce{NH3}. Another chemical pathway is for the volume species \ce{NH} and \ce{NH2} to bind to the wall midway through this process.
	
	\item In the partially ionized environment of the plasma volume, there are multiple ion species. Some of these ion species are highly reactive, and can produce \ce{NH3} in a number of different ways. For example, when \ce{H3+} encounters \ce{NH2-}, they recombine to produce \ce{NH3} (via \ce{H3+} + \ce{NH2-} $\rightarrow$ \ce{H2} + \ce{NH3}). If \ce{NH2-} encounters \ce{NH4+}, they can also produce \ce{NH3} (via \ce{NH2-} + \ce{NH4+} $\rightarrow$ \ce{NH2} + \ce{NH3} + \ce{H}). In this case, a curious fact is that \ce{NH4+} can only exist if it has been produced from \ce{NH3} (e.g.~via \ce{NH+} + \ce{NH3} $\rightarrow$ \ce{N} + \ce{NH4+}). Therefore, there is an interplay between ions and neutral species, where they produce each other recursively with the help of other species. Another example is when \ce{NH3} is produced by neutralization of \ce{NH3+}, but \ce{NH3+} itself is produced from \ce{NH3} and other ions (e.g.~via \ce{H2+} + \ce{NH3} $\rightarrow$ \ce{H2} + \ce{NH3+}). All of these mechanisms contribute to the co-existence and continuous interplay between ions and neutrals in the plasma volume.
	
	\item Besides ion reactions, \ce{NH3} can also be produced from neutrals, especially through reactions that involve \ce{NH2} (e.g.~\ce{NH} + \ce{NH2} $\rightarrow$ \ce{N} + \ce{NH3}; or \ce{2NH2} $\rightarrow$ \ce{NH} + \ce{NH3}). The heavier species \ce{N2H2}, \ce{N2H3} and \ce{N2H4}, if present, can also react with \ce{NH2} to produce \ce{NH3} (e.g.~\ce{N2H2} + \ce{NH2} $\rightarrow$ \ce{H} + \ce{N2} + \ce{NH3}). However, the origin of \ce{NH2} is difficult to disentangle, since it appears that \ce{NH2} arises from the neutralization of \ce{NH2+}, but \ce{NH2+} itself arises from the ionization of \ce{NH2} when it collides with ions such as \ce{N+}, \ce{H2+} or \ce{NH+}. Another pathway to \ce{NH3} is through neutralizing reactions involving \ce{NH2-}, but this also unclear, since \ce{NH2-} itself originates from electron impact on \ce{NH3}. In conclusion, these species (i.e.~\ce{NH2}, \ce{NH2+}, \ce{NH2-}, \ce{NH3}) seem to coexist and continuously recreate each other with the help of other species.
	
\end{itemize}

Figure~\ref{fig:scheme:graph} shows a graph representation of the chemical scheme that may shed some light on these intricate mechanisms. Here, each arrow represents a reactant-product relationship (reactant$\rightarrow$product) between two species, so a single reaction (such as \ce{H3+} + \ce{NH2-} $\rightarrow$ \ce{H2} + \ce{NH3}) may be drawn as multiple arrows (e.g.~\ce{H3+}$\rightarrow$\ce{H2}; \ce{H3+}$\rightarrow$\ce{NH3};  \ce{NH2-}$\rightarrow$\ce{H2}; \ce{NH2-}$\rightarrow$\ce{NH3}). The solid arrows represent reactions that are to be kept in the reduced chemical scheme, while the dashed arrows represent candidates for removal.

One of the main features that can be observed in this graph is the clear separation between volume and surface chemistry. It is clear that \ce{H} and \ce{N} will attach to the wall and, from there, there is a pathway towards the production of \ce{NH3} through \ce{wall\_NH2}. On the other hand, the volume chemistry shows a multitude of pathways towards \ce{NH3}, as well as the reciprocal relationships that exist among \ce{NH3}, \ce{NH2}, \ce{NH2-}, \ce{NH2+}, and even \ce{NH4+}. This visual portrayal is in agreement with our analysis above.

\section{Conclusion}
\label{sec:conclusion}

In this work, we developed a machine learning model inspired by a Petri net representation of the chemical scheme. When such representation is converted to matrix form, it becomes possible to use standard machine learning constructs. Here, we defined a learning model based on matrix-vector multiplication, and on the ReLU and softmax functions. Such model is trained by minimizing the Kullback–Leibler divergence in order to find a set of reaction weights. These weights become the basis for reducing the chemical scheme.

The model was applied to the chemistry of low-temperature \ce{N2}--\ce{H2} plasmas. Despite being a relatively simple setup, this gas discharge exhibits a rich chemistry, where both ions and neutrals, as well as volume and surface species, have an interconnected role. With the proposed model, it was possible to identify the main reactions and chemical pathways involved in the production of \ce{NH3}. In future work, we plan to further develop the approach by comparing the simulation results obtained with the full and the reduced chemical schemes, and also validate them against experimental results obtained in the lab.


\begin{credits}
	\subsubsection{\ackname} L. L. Alves would like to acknowledge the support of FCT (Fundação para a Ciência e a Tecnologia, I.P.) through project reference \href{https://doi.org/10.54499/2022.04128.PTDC}{2022.04128.PTDC}. IPFN activities were supported by FCT through project references \href{https://doi.org/10.54499/UIDB/50010/2020}{UIDB/50010/2020}, \href{https://doi.org/10.54499/UIDP/50010/2020}{UIDP/50010/2020} and \href{https://doi.org/10.54499/LA/P/0061/2020}{LA/P/0061/202}.
\end{credits}

\bibliographystyle{splncs04unsrt}
\bibliography{references}

\begin{thebibliography}{10}
\providecommand{\url}[1]{\texttt{#1}}
\providecommand{\urlprefix}{URL }
\providecommand{\doi}[1]{https://doi.org/#1}

\bibitem{Erisman_2008}
Erisman, J.W., Sutton, M.A., Galloway, J., Klimont, Z., Winiwarter, W.: How a
  century of ammonia synthesis changed the world. Nature Geoscience
  \textbf{1}(10),  636--639 (2008)

\bibitem{Carreon_2019}
Carreon, M.L.: Plasma catalytic ammonia synthesis: state of the art and future
  directions. Journal of Physics D: Applied Physics  \textbf{52}(48),  483001
  (2019)

\bibitem{Chatain_2023}
Chatain, A., Morillo-Candas, A.S., Vettier, L., Carrasco, N., Cernogora, G.,
  Guaitella, O.: Characterization of a {DC} glow discharge in \ce{N2}--\ce{H2}
  with electrical measurements and neutral and ion mass spectrometry. Plasma
  Sources Science and Technology  \textbf{32}(3),  035002 (2023)

\bibitem{Encrenaz_1974}
Encrenaz, T., Owen, T., Woodman, J.H.: The abundance of ammonia on {Jupiter},
  {Saturn} and {Titan}. Astronomy and Astrophysics  \textbf{37}(1),  49--55
  (1974)

\bibitem{Huang_2022}
Huang, J., Seager, S., Petkowski, J.J., Ranjan, S., Zhan, Z.: Assessment of
  ammonia as a biosignature gas in exoplanet atmospheres. Astrobiology
  \textbf{22}(2),  171--191 (2022)

\bibitem{Chudjak_2021}
Chudják, S., Kozáková, Z., Krčma, F.: Study of chemical processes initiated
  by electrical discharge in {Titan}-related atmosphere at laboratory
  temperature and pressure. ACS Earth and Space Chemistry  \textbf{5}(3),
  535--543 (2021)

\bibitem{Jimenez-Redondo_2020}
{Jiménez-Redondo}, M., Chatain, A., Guaitella, O., Cernogora, G., Carrasco,
  N., Alves, L.L., Marques, L.: \ce{N2}--\ce{H2} capacitively coupled
  radio-frequency discharges at low pressure: {II. Modeling} results: the
  relevance of plasma-surface interaction. Plasma Sources Science and
  Technology  \textbf{29}(8),  085023 (2020)

\bibitem{Gordiets_1998a}
Gordiets, B., Ferreira, C.M., Pinheiro, M.J., Ricard, A.: Self-consistent
  kinetic model of low-pressure \ce{N2}--\ce{H2} flowing discharges: {I.
  Volume} processes. Plasma Sources Science and Technology  \textbf{7}(3), ~363
  (1998)

\bibitem{Gordiets_1998b}
Gordiets, B., Ferreira, C.M., Pinheiro, M.J., Ricard, A.: Self-consistent
  kinetic model of low-pressure \ce{N2}--\ce{H2} flowing discharges: {II.
  Surface} processes and densities of \ce{N}, \ce{H}, \ce{NH3} species. Plasma
  Sources Science and Technology  \textbf{7}(3), ~379 (1998)

\bibitem{Marwan_2011}
Marwan, W., Wagler, A., Weismantel, R.: Petri nets as a framework for the
  reconstruction and analysis of signal transduction pathways and regulatory
  networks. Natural Computing  \textbf{10}(2),  639--654 (2011)

\bibitem{Koch_2010}
Koch, I.: Petri nets -- {A} mathematical formalism to analyze chemical reaction
  networks. Molecular Informatics  \textbf{29}(12),  838--843 (2010)

\bibitem{Asadi_2020}
Asadi, B., Jiang, H.: On approximation capabilities of {ReLU} activation and
  softmax output layer in neural networks. arXiv 2002.04060  (2020)

\bibitem{Ruder_2017}
Ruder, S.: An overview of gradient descent optimization algorithms. arXiv
  1609.04747  (2017)

\bibitem{Tejero-del-Caz_2019}
{Tejero-del-Caz}, A., Guerra, V., Gonçalves, D., da~Silva, M.L., Marques, L.,
  Pinhão, N., Pintassilgo, C.D., Alves, L.L.: The {LisbOn} {KInetics}
  {Boltzmann} solver. Plasma Sources Science and Technology  \textbf{28}(4),
  043001 (2019)

\bibitem{Guerra_2019}
Guerra, V., {Tejero-del-Caz}, A., Pintassilgo, C.D., Alves, L.L.: Modelling
  \ce{N2}--\ce{O2} plasmas: volume and surface kinetics. Plasma Sources Science
  and Technology  \textbf{28}(7),  073001 (2019)

\bibitem{Kingma_2014}
Kingma, D.P., Ba, J.: Adam: A method for stochastic optimization.
  arXiv:1412.6980  (2014)

\end{thebibliography}

\end{document}